\documentclass{cjaa}

\usepackage{graphicx}
\input{epsf.sty}
\input{psfig.sty}

\begin{document}

   \title{The amplitude of mass fluctuations and mass density of
   the Universe constrained by strong gravitational lensing}

\author{Da-Ming Chen
\inst{}}
\institute{National Astronomical Observatories, Chinese
Academy of Sciences, Beijing 100012, China\\
\email{cdm@bao.ac.cn} }

  \date{Received~~2003 month day; accepted~~2003~~month day}
   \abstract{We investigate the linear amplitude of mass
   fluctuations in the universe, $\sigma_8$, and the present mass
   density parameter of the Universe, $\Omega_\mathrm{m}$, from
   the statistical strong gravitational lensing. We use the two
   populations of lens halos model with fixed cooling mass scale
   $M_\mathrm{c}=3\times 10^{13}h^{-1}M_{\sun}$ to match the
   observed lensing probabilities, and leave $\sigma_8$ or
   $\Omega_\mathrm{m}$ as a free parameter to be constrained by
   data. Another varying parameter is the equation of state of dark energy
   $\omega$, and its typical values of $-1$, $-2/3$, $-1/2$ and $-1/3$
   are investigated.
   We find that $\sigma_8$ is degenerate with $\Omega_\mathrm{m}$
   in a way similar to that suggested by present day cluster
   abundance as well as cosmic shear lensing measurements:
   $\sigma_8\Omega_\mathrm{m}^{0.6}\approx 0.33$
   (Bahcall \& Bode~\cite{bahcall03a} and
   references therein). However, both $\sigma_8\leq 0.7$ and
   $\Omega_\mathrm{m}\leq 0.2$
   can be safely ruled out, the best value is when $\sigma_8=1.0$,
   $\Omega_\mathrm{m}=0.3$ and
   $\omega=-1$. This result is different from that obtained by
   Bahcall \& Bode (\cite{bahcall03a}), who gives
   $\sigma_8 =0.98\pm 0.1$ and
$\Omega_m =0.17\pm 0.05$.
    For $\sigma_8=1.0$, higher value of
   $\Omega_\mathrm{m}=0.35$ requires $\omega=-2/3$ and
   $\Omega_\mathrm{m}=0.40$ requires $\omega=-1/2$.
   }

\authorrunning{D. -M. Chen}
\titlerunning{$\sigma_8$ and mass density of the Universe}

\maketitle

\section{Introduction}
\label{sect:intro}
The amplitude of mass fluctuations, denoted as $\sigma_8$ when
referring to the rms linear density fluctuation in spheres of
radius $8h^{-1}$Mpc at $z=0$, is a fundamental cosmological
parameter that describes the normalization of the linear spectrum
of mass fluctuations in the early universe. Assuming Gaussian
initial fluctuations, the evolution of structure in the universe
depends exponentially on this parameter (for an excellent review
see Bahcall \& Bode~\cite{bahcall03a}).

Recent observations suggest an amplitude that ranges in value from
$\sigma_8 \sim 0.7$ to a high value of $\sigma_8 \sim 1.1$. The
low amplitude values of $\sigma_8 \sim 0.7$ are suggested by
current observations of the cosmic microwave background (CMB)
spectrum of fluctuations (Netterfield et al.~\cite{nette}; Sievers
et al.~\cite{sieve}; Bond et al.~\cite{bond}; Ruhl et
al.~\cite{ruhle}) and by recent observations of the present
cluster abundance as well as cosmic shear lensing measurements
(Jarvis et al.~\cite{jarvi}; Hamana et al.~\cite{haman};
Seljak~\cite{selja}). However, these determinations of $\sigma_8$
are degenerate with other parameters like mass density parameter
$\Omega_\mathrm{m}$.  The evolution of cluster abundance with
time, especially for the most massive clusters, breaks the
degeneracy between $\sigma_8$ and $\Omega_m$ (e.g., Peebles, Daly
\& Juszkiewicz~\cite{PDJ89}; Eke, Cole \& Frenk~\cite{ECF96};
Oukbir \& Blanchard~\cite{OB97}; Bahcall, Fan \&
Cen~\cite{BFC97};Carlberg et al.~\cite{CMYE97}; Bahcall \&
Fan~\cite{bah98}; Donahue \& Voit~\cite{DV99};
Henry~\cite{Henry00}). This evolution depends strongly on
$\sigma_8$, and only weakly on $\Omega_m$ or other parameters.
Bahcall \& Bode (\cite{bahcall03a}) used the abundance of the most
massive clusters observed at $z\sim 0.5-0.8$ to place a strong
limit on $\sigma_8$ and found that $\sigma_8 =0.98\pm 0.1$,
$\Omega_m =0.17\pm 0.05$, and low $\sigma_8$ values ($\la 0.7$)
are unlikely. In the model of one population of halos
(Navarro-Frenk-White, NFW) combined with each galactic halo a
central point mass, the lensing probabilities are shown to be
sensitive to $\sigma_8$ (chen~\cite{chen03a}, hereafter, paper I).

 In this paper, we use
the model of the two populations of lens halos to calculate the
lensing probabilities in flat quintessence cold dark matter (QCDM)
cosmology with different cosmic equations of state $\omega$
(Chen~\cite{chen03b}, hereafter, paper II), leaving $\sigma_8$ and
$\Omega_\mathrm{m}$ as free parameters to be constrained from the
Jodrell-Bank VLA Astrometric Survey (JVAS) and the Cosmic Lens
All-Sky Survey (CLASS; Browne et al.~\cite{browne};
Helbig~\cite{helbig}; Browne et al.~\cite{browne02}; Myers et
al.~\cite{myers}).

\section{Lensing probabilities}
When the quasars at the mean redshift $<z_{\mathrm{s}}>=1.27$ are
lensed by foreground CDM halos of galaxies and clusters of
galaxies, the lensing probability with image separations larger
than $\Delta\theta$ and flux density ratio less than
$q_{\mathrm{r}}$ is (Schneider et al. \cite{schne})
\begin{equation}
P(>\Delta\theta, <q_{\mathrm{r}})=
\int^{z_{\mathrm{s}}}_0\frac{dD_{\mathrm{L}}(z)}
{dz}dz\int^{\infty}_0\bar{n}(M,z)\sigma(M,z)B(M,z)dM,
\label{prob1}
\end{equation}
where $D_{\mathrm{L}}(z)$ is the proper distance from the observer
to the lens located at redshift $z$. And $\bar{n}(M,z)$ is the
physical number density of virialized dark halos of masses between
$M$ and $M+dM$ at redshift $z$ given by Jenkins et al.
(\cite{jenki}). The cross section $\sigma(M,z)$ is mass and
redshift dependent, and is sensitive to flux density ratio of
multiple images $q_\mathrm{r}$ for SIS halos,
\begin{equation}
\sigma(M,z)=\pi\xi_0^2\vartheta(M-M_{\mathrm{min}})\times \cases{
y_{\mathrm{cr}}^2,&for $\Delta\theta\leq\Delta\theta_0$; \cr
y_{\mathrm{cr}}^2-y_{\Delta\theta}^2,&for
$\Delta\theta_0\leq\Delta\theta<\Delta\theta_{y_{\mathrm{cr}}}$;\cr
0,&for $\Delta\theta\geq\Delta\theta_{y_{\mathrm{cr}}}$,\cr}
\label{cross}
\end{equation}
where $\vartheta(x)$ is a step function, and $M_{\mathrm{min}}$ is
the minimum mass of halos above which lenses can produce images
with separations greater than $\Delta\theta$. It is shown (in
paper II) that the contributions from galactic central
supermassive black holes can be ignored when $q_{\mathrm{r}}\leq
10$, so the lensing equation for SIS halos is simply $y=x-|x|/x$,
where $x=|\vec{x}|$ and $y=|\vec{y}|$, which
 are related to the position vector in the lens plane and source
plane as $\vec{\xi}=\vec{x}\xi_0$ and $\vec{\eta}=\vec{y}\eta_0$,
respectively. The length scales in the lens plane and the source
plane are defined to be
$\xi_0=4\pi(\sigma_{v}/c)^2(D_\mathrm{L}^\mathrm{A}
D_\mathrm{LS}^\mathrm{A})/D_\mathrm{S}^\mathrm{A}$ and
$\eta_0=\xi_0D_\mathrm{S}^\mathrm{A}/D_\mathrm{L}^\mathrm{A}$.
 Since the surface mass density is circularly symmetric, we can
extend both $x$ and $y$ to their opposite values in our actual
calculations for convenience. From the lensing equation, an image
separation for any $y$ can be expressed as
$\Delta\theta(y)=\xi_0\Delta x(y)/D_\mathrm{L}^\mathrm{A}$, where
$\Delta x(y)$ is the image separation in lens plane for a given
$y$. So in Eq.(\ref{cross}), the source position
$y_{\Delta\theta}$, at which a lens produces the image separation
$\Delta\theta$, is the reverse of this expression. And
$\Delta\theta_0=\Delta\theta(0)$ is the separation of the two
images which are just on the Einstein ring;
$\Delta\theta_{y_{\mathrm{cr}}}=\Delta\theta(y_{\mathrm{cr}})$ is
the upper-limit of the separation above which the flux ratio of
the two images will be greater than $q_{\mathrm{r}}$. Note that
since $M_{\mathrm{DM}}$ is related to $\Delta\theta$ through
$\xi_0$ and $\sigma_v^2=GM_\mathrm{DM}/2r_\mathrm{vir}$, we can
formally write $M_{\mathrm{DM}}=M_{\mathrm{DM}}(\Delta\theta(y))$
and determine $M_{\mathrm{min}}$ for galaxy-size lenses by
$M_{\mathrm{min}}=M_{\mathrm{DM}}(\Delta\theta(y_{\mathrm{cr}}))$.

According to the model of two populations of halos, cluster-size
halos are modeled as NFW profile:
$\rho_\mathrm{NFW}=\rho_\mathrm{s}r_\mathrm{s}^3/
[r(r+r_\mathrm{s})^2]$ , where $\rho_\mathrm{s}$ and
$r_\mathrm{s}$ are constants. We can define the mass of a halo to
be the mass within the virial radius of the halo $r_\mathrm{ vir}$
: $M_\mathrm{DM}=4\pi\rho_\mathrm{s}r_\mathrm{s}^3f(c_1)$, where
$f(c_1)=\ln(1+c_1)-c_1/(1+c_1)$, and $c_1=r_\mathrm{
vir}/r_\mathrm{s}=9(1+z)^{-1}(M/1.5\times
10^{13}h^{-1}M_{\sun})^{-0.13}$ is the concentration parameter,
for which we have used the fitting formula given by Bullock et al.
(\cite{bullo}). The lensing equation for NFW lenses is as usual
$y=x-\mu_s g(x)/x$ (Li \& Ostriker~\cite{li}), where
$y=|\vec{y}|$,
$\vec{\eta}=\vec{y}D_\mathrm{S}^\mathrm{A}/D^\mathrm{A}_\mathrm{L}$
is the position vector in the source plane, in which
$D_\mathrm{S}^\mathrm{A}$ and $D_\mathrm{L}^\mathrm{A}$ are
angular-diameter distances from the observer to the source and to
the lens respectively.  $x=|\vec{x}|$ and
$\vec{x}=\vec{\xi}/r_\mathrm{s}$, $\vec{\xi}$ is the position
vector in the lens plane.  The parameter
$\mu_\mathrm{s}=4\rho_\mathrm{s}r_\mathrm{s}/\Sigma_\mathrm{cr}$
is $x$ independent, in which $\Sigma_\mathrm{cr}=(c^2/4\pi
G)(D_\mathrm{S}^\mathrm{A}/D_\mathrm{L}^\mathrm{A}
D_\mathrm{LS}^\mathrm{A})$ is critical surface mass density, with
$c$ the speed of light, $G$ the gravitational constant and
$D_\mathrm{LS}^\mathrm{A}$ the angular-diameter distance from the
lens to the source. The function $g(x)$ has a analytical
expression originally given by Bartelmann (\cite{barte}). The
cross section for the cluster-size NFW lenses is well studied (Li
\& Ostriker~\cite{li}). The lensing equation is $y=x-\mu_s
g(x)/x$, and the multiple images can be produced only if $|y|\leq
y_{\mathrm{cr}}$, where $y_{\mathrm{cr}}$ is the maximum value of
$y$ when $x<0$, which is determined by $dy/dx=0$, and the cross
section in the lens plane is simply $\sigma(M, z)=\pi
y_\mathrm{cr}^2r_\mathrm{s}^2$.

 As for the
magnification bias $B(M,z)$, we use the result given by Li \&
Ostriker (\cite{li}) for NFW lenses. For singular isothermal
sphere (SIS) model, the magnification bias is
$B_\mathrm{SIS}\approx 4.76$.

We consider In this paper the spatially flat QCDM cosmology
models. The density parameter $\Omega_\mathrm{m}$ ranges from
$0.2$ to $0.4$ as suggested by all kinds of measurements (e.g.,
Peebles \& Ratra~\cite{peebles03} and the references therein). We
investigate the varying parameter $\sigma_8$ within its entire
observational range from 0.7 to 1.1 (e.g., Bahcall \&
Bode~\cite{bahcall03a}). The Hubble parameter is $h=0.75$. Three
negative values of $\omega$ in equation of state
$p_\mathrm{Q}=\omega\rho_\mathrm{Q}$, with $\omega=-1$
(cosmological constant), $\omega=-2/3$, $\omega=-1/2$ and
$\omega=-1/3$  are investigated. We use the conventional form to
express the redshift $z$ dependent linear power spectrum for the
matter density perturbation and the linear growth suppression
factor of the density field in QCDM cosmology established by Ma et
al. (\cite{ma1999}), which are needed in Eq. (\ref{prob1}).

\section{Discussion and conclusions}

Since the lensing rate is sensitive to the source redshift
$z_\mathrm{s}$, results can be affected considerably by including
the redshift distribution into calculations (e.g., Sarbu, Rusin \&
Ma~\cite{sarbu}). However, since its distribution in the
JVAS/CLASS survey is still poorly understood, we use the estimated
mean value of $<z_\mathrm{s}>=1.27$ (Marlow et al.~\cite{marlow};
Chae et al.~\cite{chae}; Oguri~\cite{oguri}; Huterer \&
Ma~\cite{huter}; Paper I).

\begin{figure}
\resizebox{\hsize}{!}{\includegraphics{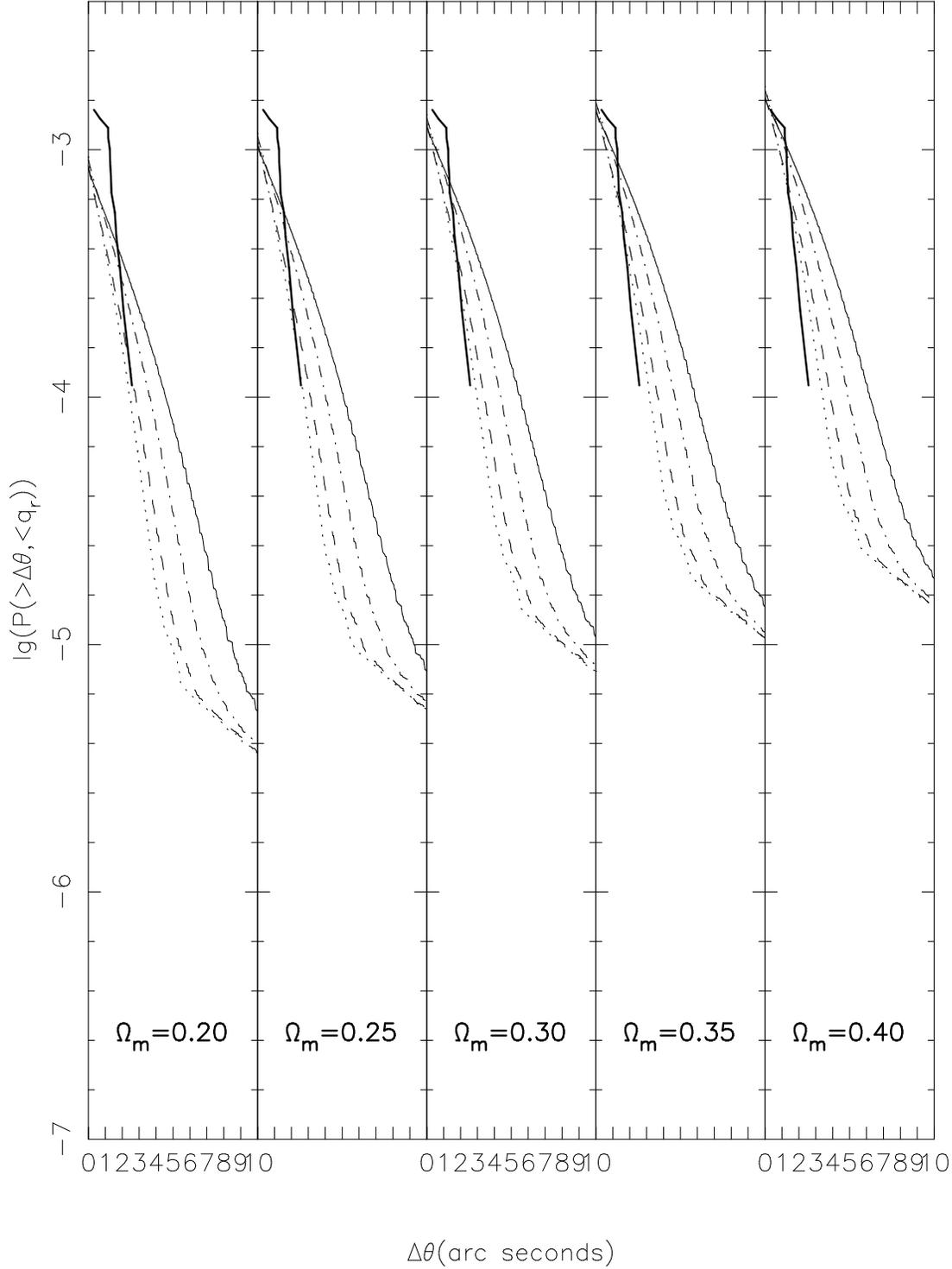}}
   \caption{Predicted lensing
probability with image separations $>\Delta\theta$ and flux
density ratios $<q_{\mathrm{r}}$ in $\Lambda$CDM cosmology. The
cluster-size lens halos are modelled by the NFW profile, and
galaxy-size lens halos by NFW+BULGE. Instead of SIS, we treat the
bulge as a point mass, its value $M_{\mathrm{eff}}$ is so selected
for each $q_{\mathrm{r}}$ that the predicted lensing probability
can match the results of JVAS/CLASS represented by histogram. In
each panel, the solid, dash-dotted, dashed and dotted lines (from
top downwards) stand for, respectively, the matched values of the
pair $(q_\mathrm{r}, M_\mathrm{eff}/M_\bullet)$ ($M_\bullet$ is a
galactic central black hole mass) of $(10, 200)$, $(100, 100)$,
$(1000, 50)$ and $(10000, 30)$.
   $<z_\mathrm{s}>=1.27$ and $\sigma_8=1.0$ for all panels here, and from left to right,
   $\Omega_\mathrm{m}$ is 0.2, 0.25, 0.3, 0.35 and 0.4, respectively}
   \label{fig1}
\end{figure}

For comparison, we plot in Fig.~\ref{fig1} the lensing probability
versus image separation angle for each set of parameters of
cosmology and lens halo models with the same values as those taken
in the right panel of Fig. 1 in paper I, except the mean value of
the redshift of quasars $<z_\mathrm{s}>$, the amplitude of mass
fluctuations $\sigma_8$ and the mass density parameter
$\Omega_\mathrm{m}$. This is a model of one population of halos
(NFW) combined with each galactic halo a central point mass
($M_\mathrm{eff}$). Other than a higher value of
$<z_\mathrm{s}>=1.5$ used in paper I, we use an estimated value of
$<z_\mathrm{s}>=1.27$ in this paper. We use a slightly higher
value of $\sigma_8=1.0$ (while in paper I, this value is
$\sigma_8=0.95$). The histogram represents the results of
JVAS/CLASS; the solid, dash-dotted, dashed and dotted lines (from
top downwards) stand for, respectively, the matched values of the
pair $(q_\mathrm{r}, M_\mathrm{eff}/M_\bullet)$ ($M_\bullet$ is a
galactic central black hole mass) of $(10, 200)$, $(100, 100)$,
$(1000, 50)$ and $(10000, 30)$. Five values of $\Omega_\mathrm{m}$
ranging from 0.2 to 0.4 (as explicitly indicated in each panel)
are chosen to see its effect on lensing probabilities. We find
that lensing probability is sensitive to $\Omega_\mathrm{m}$,
however, the best fit parameters are only when $\omega=-1$ and
$\Omega_\mathrm{m}=0.4$, which is different from the result given
in paper I. The reason is that the lensing probabilities are quite
sensitive to the mean redshift of quasars $<z_\mathrm{s}>$, the
higher redshift will produce a larger value of lensing
probability. This means that the NFW+point-mass model for galaxy
size lens halos indeed reduces the probabilities considerably when
small image flux density ratio is taken into account, which can be
confirmed when compared with the SIS model and the discussion
below.

As just mentioned, when we use NFW+point-mass to model galaxy size
lens halos, the predicted lensing probabilities can match
observations only when a higher value of $<z_\mathrm{s}>$ is used.
We pointed out in paper II that a two populations of lens halos
model with mass distributions NFW ($M_\mathrm{DM}>M_\mathrm{c}$)
and SIS ($M_\mathrm{DM}<M_\mathrm{c}$), can match observations
better, even when a reasonable lower value of $<z_\mathrm{s}>$ and
$M_\mathrm{c}$ are used. We chose the cooling mass scale to be
$M_\mathrm{c}=3.0\times 10^{13}h^{-1}M_{\sun}$ in this paper
rather than $M_\mathrm{c}=5.0\times 10^{13}h^{-1}M_{\sun}$ used in
paper I.

\begin{figure}
\resizebox{\hsize}{!}{\includegraphics{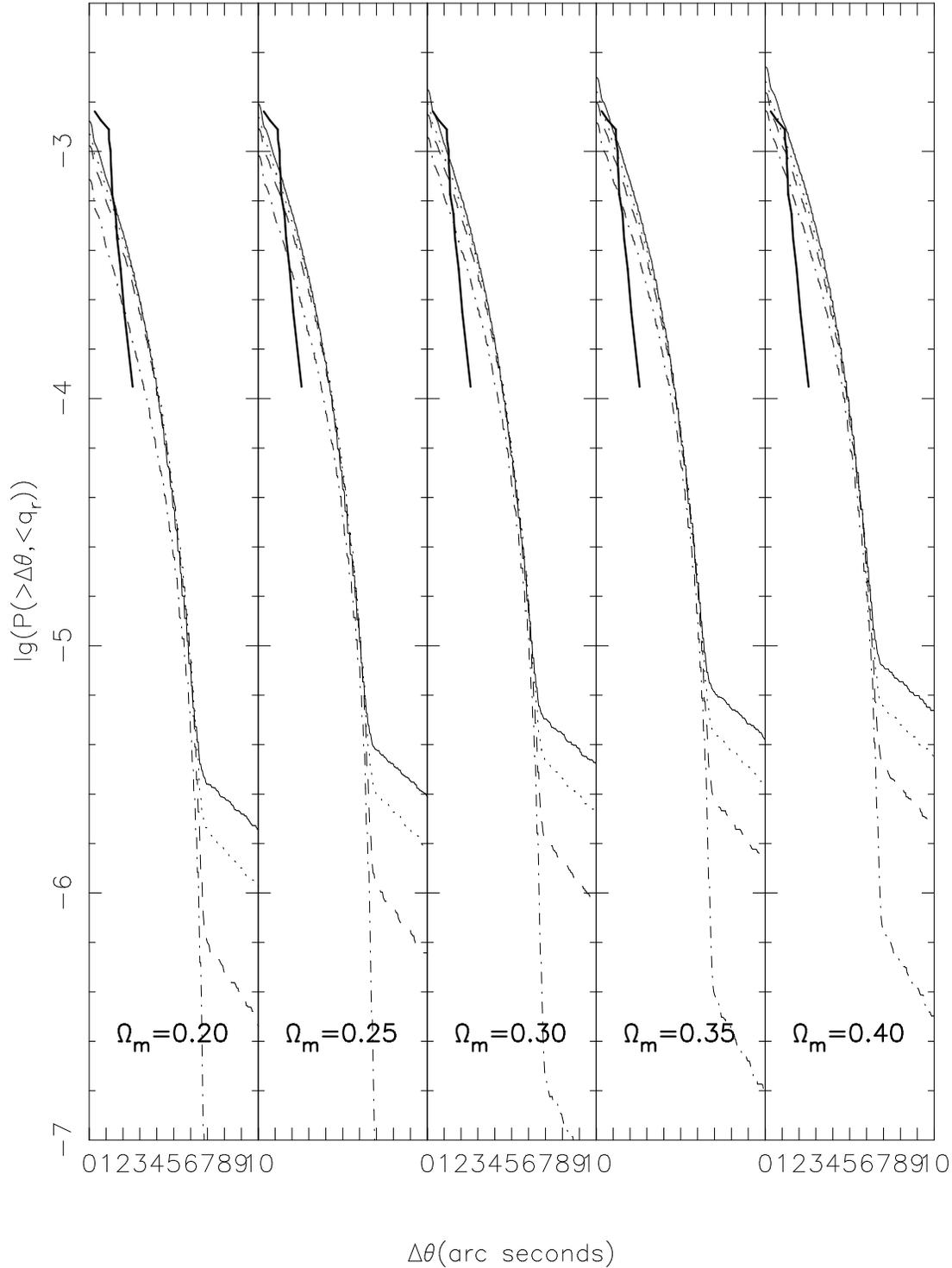}}
   \caption{The
integral lensing probabilities with image separations larger than
$\Delta\theta$ and flux density ratio less than $q_{\mathrm{r}}$,
for quasars at mean redshift $<z_\mathrm{s}>=1.27$ lensed by NFW
($M_\mathrm{DM}>M_\mathrm{c}$) and SIS
($M_\mathrm{DM}<M_\mathrm{c}$) halos. In each panel,
   $q_\mathrm{r}=10.0$ and
   $\sigma_8=1.0$, and  from left to right,
   $\Omega_\mathrm{m}$ is 0.2, 0.25, 0.3, 0.35 and 0.4, respectively. The solid, dashed,
   dash-dotted and dotted lines stand for $\omega=-1$, -2/3, -1/2  and -1/3, respectively.}
   \label{fig2}
\end{figure}

So it would be interesting to investigate both the
$\Omega_\mathrm{m}$ and $\sigma_8$ dependent lensing probability
with the combined SIS and NFW model. In each panel of
fig.~\ref{fig2}, the parameters are: $q_\mathrm{r}=10$,
$\sigma_8=1.0$. And $\Omega_\mathrm{m}$ takes five values as in
fig.~\ref{fig1} (as explicitly indicated). We find that, lensing
probability is also sensitive to $\Omega_\mathrm{m}$, and clealy,
$\Omega_\mathrm{m}=0.2$ can be safely ruled out. For $\omega=-1$
(cosmological constant), the best fit value of mass density
parameter is $\Omega_\mathrm{m}=0.3$. This result is different
from those obtained with other methods (see Bahcall \&
Bode~\cite{bahcall03a}). Higher value of
   $\Omega_\mathrm{m}=0.35$ requires $\omega=-2/3$ and
   $\Omega_\mathrm{m}=0.40$ requires $\omega=-1/2$.

\begin{figure}
\resizebox{\hsize}{!}{\includegraphics{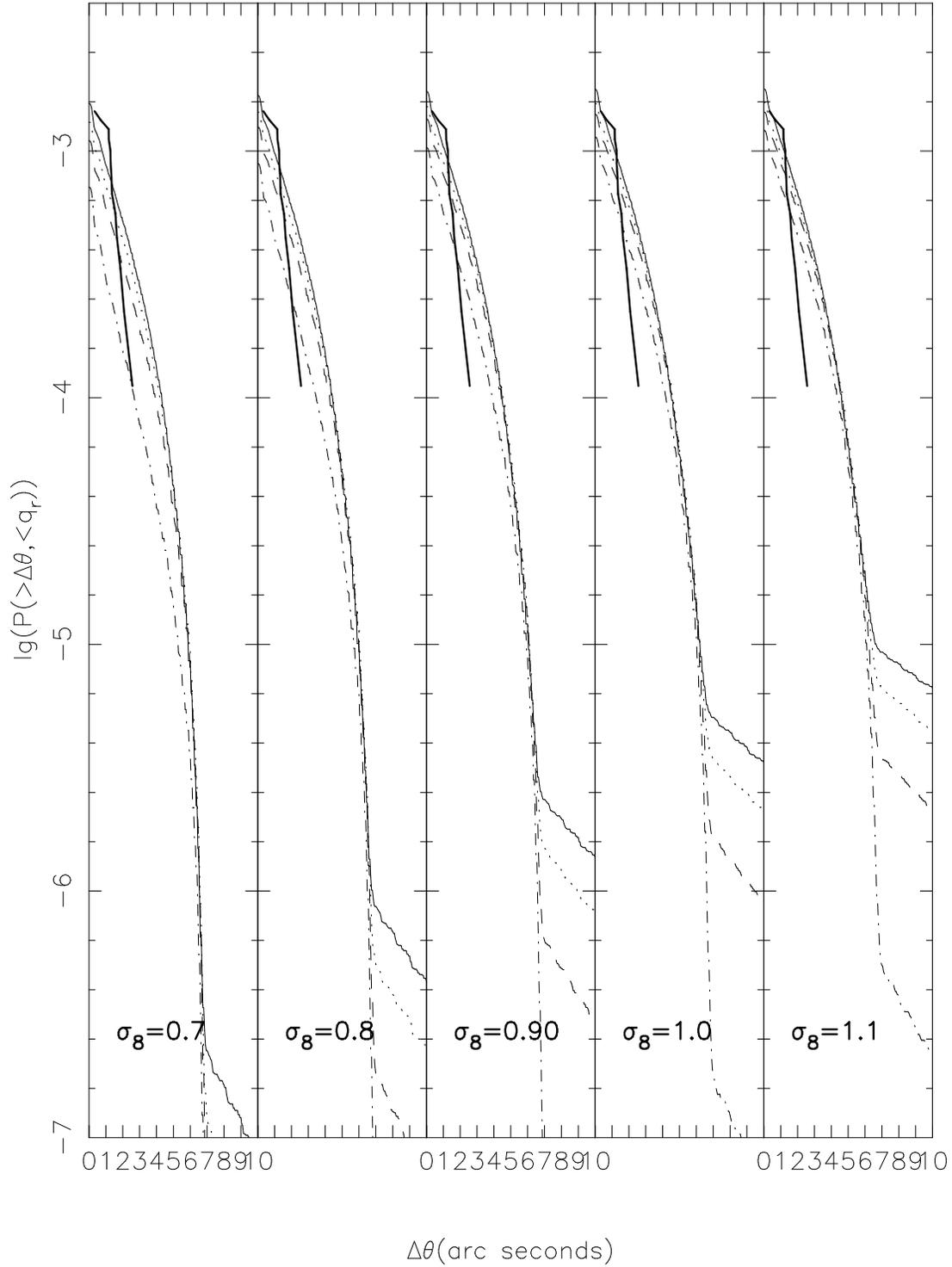}}
   \caption{Same as Fig. 2, except
   the value of $\sigma_8$, the value of which, from left to right,
   is 0.7, 0.8, 0.9, 1.0 and 1.1, respectively. In each panel, $\Omega_\mathrm{m}=0.3$.}
   \label{fig3}
\end{figure}

\begin{figure}
\resizebox{\hsize}{!}{\includegraphics{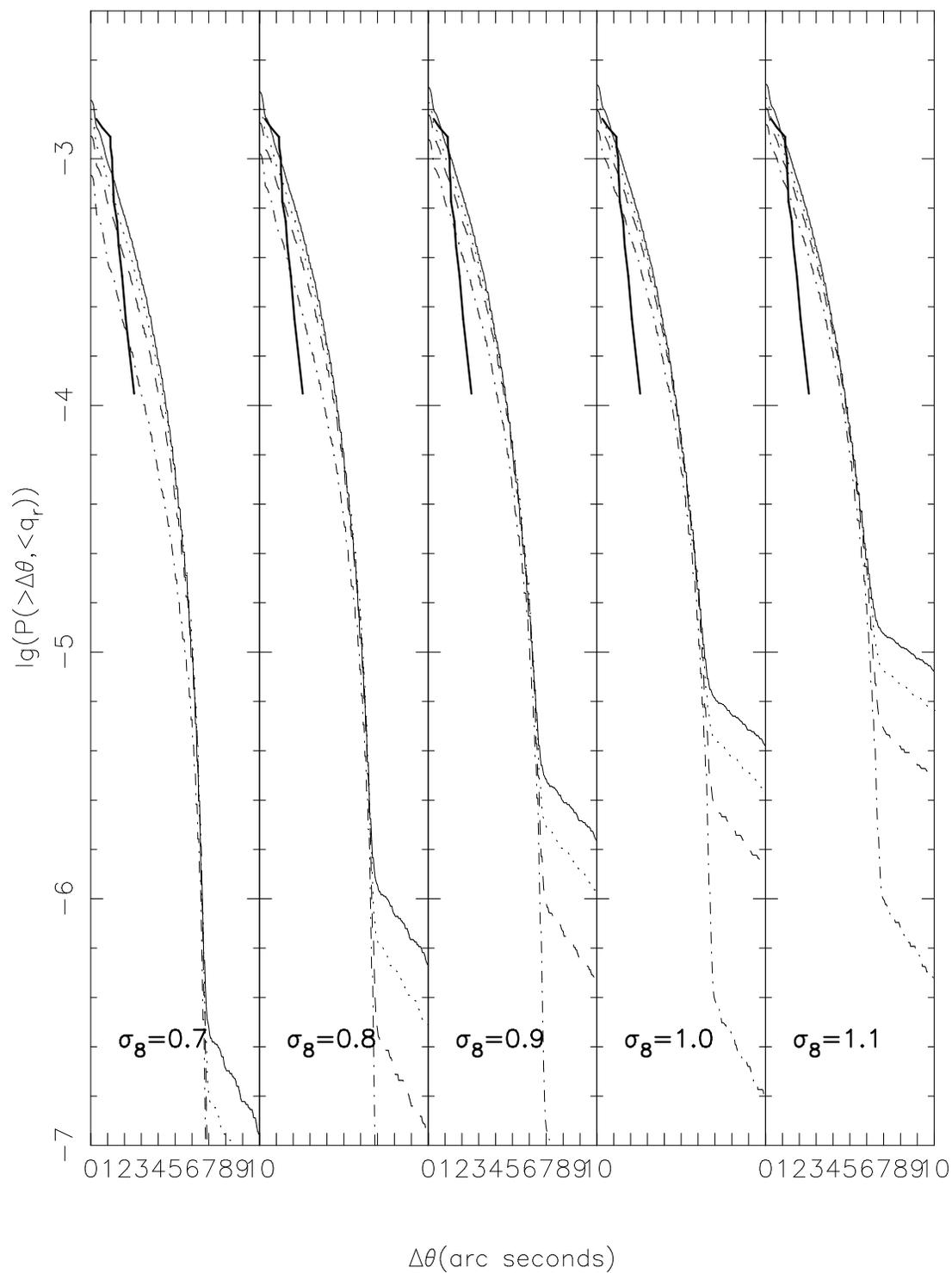}}
   \caption{Same as Fig. 3, except that
   $\Omega_\mathrm{m}=0.35$.}
   \label{fig4}
\end{figure}

Our model prefers a higher value of $\Omega_\mathrm{m}\geq 0.3$.
In Fig.~\ref{fig2}, we have already used higher values of
$\sigma_8$ ($=1.0$) and $M_\mathrm{c}$ ($=3.0\times
10^{13}h^{-1}M_{\sun}$). Lower values of $\Omega_\mathrm{m}$
require more higher values of these two parameters, which would be
out of the range suggested by other measurements. In order to see
the effect of $\sigma_8$ on lensing probability, we thus fix the
value of $\Omega_\mathrm{m}$ to be 0.3 and 0.35, respectively, and
vary $\sigma_8$ from 0.7 to 1.1 in each case. The results are
shown in Fig.~\ref{fig3} and Fig.~\ref{fig4}. We find that, when
$\Omega_\mathrm{m}=0.3$, lensing probabilities are only slightly
sensitive to $\sigma_8$ at small image separations
($0.3''<\Delta\theta<3''$), where JVAS/CLASS survey has a
well-defined sample suitable for analysis of the lens statistics.
The lensing probabilities are sensitive to $\sigma_8$ at larger
image separations, but no sample suitable for analysis exists in
this range. While $\sigma_8\leq 0.7$ seems unlikely, all the
values in the range $0.8\leq\sigma_8\leq 1.1$ are possible. For a
larger value of $\Omega_\mathrm{m}=0.35$, as shown in
Fig.~\ref{fig4}, the lensing probabilities are more sensitive to
$\sigma_8$ than in Fig.~\ref{fig3}. In this case, even
$\sigma_8=0.7$ is acceptable (it predicts 12.5 lenses with image
separation $\geq 0.3''$, while the observed value is 13), and
$\sigma_8=1.1$ matches a value of equation of state of dark energy
to be $\omega=-2/3$.

Note that Chae et al. (\cite{chae}) reported the main results on
cosmological parameters (matter density $\Omega_\mathrm{m}$ and
equation of state for dark energy $\omega$) from a likelihood
analysis of lens statistics, they gave
$\Omega_\mathrm{m}=0.31^{+0.27}_{-0.14}$ and
$\omega=0.55^{+0.18}_{-0.11}$, both at $68\%$ confidence level.
Our results, although not precise, are in agreement with theirs.
However, Since Chae et al. (\cite{chae}) used the Schechter
luminosity function rather than Press-Schechter mass function to
account for the mass distribution, they didn't refer to
$\sigma_8$. Precise results using the same model of this paper
from a likelihood analysis will be presented in another paper.

\begin{acknowledgements}
The author thanks the anonymous referee for useful comments and
constructive suggestions. This work was supported by the National
Natural Science Foundation of China under grant No.10233040.
\end{acknowledgements}

\end{document}